\documentclass[aps,prl,twocolumn,showpacs,groupedaddress,floatfix]{revtex4}  
\usepackage{graphicx}  
\usepackage{dcolumn}   
\usepackage{bm}        
\usepackage{amssymb}   
\def\gKK{\mbox{$G^{(1)}$}}
\def\kMPl{\mbox{$k/\overline M_{\rm Pl}$}}
\def\ipb{\mbox{pb$^{-1}$}}
\newcommand{\ltsim}{\protect\raisebox{-0.7ex}{$\:\stackrel{\textstyle <}
	{\sim}\:$}}
\begin{document}



\title{Search for Randall-Sundrum Gravitons in Dilepton and Diphoton Final States}

%
\author{                                                                      
V.M.~Abazov,$^{35}$                                                           
B.~Abbott,$^{72}$                                                             
M.~Abolins,$^{63}$                                                            
B.S.~Acharya,$^{29}$                                                          
M.~Adams,$^{50}$                                                              
T.~Adams,$^{48}$                                                              
M.~Agelou,$^{18}$                                                             
J.-L.~Agram,$^{19}$                                                           
S.H.~Ahn,$^{31}$                                                              
M.~Ahsan,$^{57}$
% ...
\\                                                                            
\vskip 0.30cm                                                                 
\centerline{(D\O\ Collaboration)}                                             
\vskip 0.30cm                                                                 
}                                                                             
\affiliation{                                                                 
\centerline{$^{1}$Universidad de Buenos Aires, Buenos Aires, Argentina}       
\centerline{$^{2}$LAFEX, Centro Brasileiro de Pesquisas F{\'\i}sicas,         
                  Rio de Janeiro, Brazil}                                     
\centerline{$^{3}$Universidade do Estado do Rio de Janeiro,                   
                  Rio de Janeiro, Brazil}                                     
\centerline{$^{4}$Instituto de F\'{\i}sica Te\'orica, Universidade            
                  Estadual Paulista, S\~ao Paulo, Brazil}                     
\centerline{$^{5}$University of Alberta, Edmonton, Alberta, Canada,           
               Simon Fraser University, Burnaby, British Columbia, Canada,}   
\centerline{York University, Toronto, Ontario, Canada, and                    
         McGill University, Montreal, Quebec, Canada}                         
\centerline{$^{6}$Institute of High Energy Physics, Beijing,                  
                  People's Republic of China}                                 
\centerline{$^{7}$University of Science and Technology of China, Hefei,       
                  People's Republic of China}                                 
\centerline{$^{8}$Universidad de los Andes, Bogot\'{a}, Colombia}             
\centerline{$^{9}$Center for Particle Physics, Charles University,            
                  Prague, Czech Republic}                                     
\centerline{$^{10}$Czech Technical University, Prague, Czech Republic}        
\centerline{$^{11}$Institute of Physics, Academy of Sciences, Center          
                  for Particle Physics, Prague, Czech Republic}               
\centerline{$^{12}$Universidad San Francisco de Quito, Quito, Ecuador}        
\centerline{$^{13}$Laboratoire de Physique Corpusculaire, IN2P3-CNRS,         
                 Universit\'e Blaise Pascal, Clermont-Ferrand, France}        
\centerline{$^{14}$Laboratoire de Physique Subatomique et de Cosmologie,      
                  IN2P3-CNRS, Universite de Grenoble 1, Grenoble, France}     
\centerline{$^{15}$CPPM, IN2P3-CNRS, Universit\'e de la M\'editerran\'ee,     
                  Marseille, France}                                          
\centerline{$^{16}$Laboratoire de l'Acc\'el\'erateur Lin\'eaire,              
                  IN2P3-CNRS, Orsay, France}                                  
\centerline{$^{17}$LPNHE, IN2P3-CNRS, Universit\'es Paris VI and VII,         
                  Paris, France}                                              
\centerline{$^{18}$DAPNIA/Service de Physique des Particules, CEA, Saclay,    
                  France}                                                     
\centerline{$^{19}$IReS, IN2P3-CNRS, Universit\'e Louis Pasteur, Strasbourg,  
                France, and Universit\'e de Haute Alsace, Mulhouse, France}   
\centerline{$^{20}$Institut de Physique Nucl\'eaire de Lyon, IN2P3-CNRS,      
                   Universit\'e Claude Bernard, Villeurbanne, France}         
\centerline{$^{21}$III. Physikalisches Institut A, RWTH Aachen,               
                   Aachen, Germany}                                           
\centerline{$^{22}$Physikalisches Institut, Universit{\"a}t Bonn,             
                  Bonn, Germany}                                              
\centerline{$^{23}$Physikalisches Institut, Universit{\"a}t Freiburg,         
                  Freiburg, Germany}                                          
\centerline{$^{24}$Institut f{\"u}r Physik, Universit{\"a}t Mainz,            
                  Mainz, Germany}                                             
\centerline{$^{25}$Ludwig-Maximilians-Universit{\"a}t M{\"u}nchen,            
                   M{\"u}nchen, Germany}                                      
\centerline{$^{26}$Fachbereich Physik, University of Wuppertal,               
                   Wuppertal, Germany}                                        
\centerline{$^{27}$Panjab University, Chandigarh, India}                      
\centerline{$^{28}$Delhi University, Delhi, India}                            
\centerline{$^{29}$Tata Institute of Fundamental Research, Mumbai, India}     
\centerline{$^{30}$University College Dublin, Dublin, Ireland}                
\centerline{$^{31}$Korea Detector Laboratory, Korea University,               
                   Seoul, Korea}                                              
\centerline{$^{32}$CINVESTAV, Mexico City, Mexico}                            
\centerline{$^{33}$FOM-Institute NIKHEF and University of                     
                  Amsterdam/NIKHEF, Amsterdam, The Netherlands}               
\centerline{$^{34}$Radboud University Nijmegen/NIKHEF, Nijmegen, The          
                  Netherlands}                                                
\centerline{$^{35}$Joint Institute for Nuclear Research, Dubna, Russia}       
\centerline{$^{36}$Institute for Theoretical and Experimental Physics,        
                  Moscow, Russia}                                             
\centerline{$^{37}$Moscow State University, Moscow, Russia}                   
\centerline{$^{38}$Institute for High Energy Physics, Protvino, Russia}       
\centerline{$^{39}$Petersburg Nuclear Physics Institute,                      
                   St. Petersburg, Russia}                                    
\centerline{$^{40}$Lund University, Lund, Sweden, Royal Institute of          
                   Technology and Stockholm University, Stockholm,            
                   Sweden, and}                                               
\centerline{Uppsala University, Uppsala, Sweden}                              
\centerline{$^{41}$Lancaster University, Lancaster, United Kingdom}           
\centerline{$^{42}$Imperial College, London, United Kingdom}                  
\centerline{$^{43}$University of Manchester, Manchester, United Kingdom}      
\centerline{$^{44}$University of Arizona, Tucson, Arizona 85721, USA}         
\centerline{$^{45}$Lawrence Berkeley National Laboratory and University of    
                  California, Berkeley, California 94720, USA}                
\centerline{$^{46}$California State University, Fresno, California 93740, USA}
\centerline{$^{47}$University of California, Riverside, California 92521, USA}
\centerline{$^{48}$Florida State University, Tallahassee, Florida 32306, USA} 
\centerline{$^{49}$Fermi National Accelerator Laboratory, Batavia,            
                   Illinois 60510, USA}                                       
\centerline{$^{50}$University of Illinois at Chicago, Chicago,                
                   Illinois 60607, USA}                                       
\centerline{$^{51}$Northern Illinois University, DeKalb, Illinois 60115, USA} 
\centerline{$^{52}$Northwestern University, Evanston, Illinois 60208, USA}    
\centerline{$^{53}$Indiana University, Bloomington, Indiana 47405, USA}       
\centerline{$^{54}$University of Notre Dame, Notre Dame, Indiana 46556, USA}  
\centerline{$^{55}$Iowa State University, Ames, Iowa 50011, USA}              
\centerline{$^{56}$University of Kansas, Lawrence, Kansas 66045, USA}         
\centerline{$^{57}$Kansas State University, Manhattan, Kansas 66506, USA}     
\centerline{$^{58}$Louisiana Tech University, Ruston, Louisiana 71272, USA}   
\centerline{$^{59}$University of Maryland, College Park, Maryland 20742, USA} 
\centerline{$^{60}$Boston University, Boston, Massachusetts 02215, USA}       
\centerline{$^{61}$Northeastern University, Boston, Massachusetts 02115, USA} 
\centerline{$^{62}$University of Michigan, Ann Arbor, Michigan 48109, USA}    
\centerline{$^{63}$Michigan State University, East Lansing, Michigan 48824,   
                   USA}                                                       
\centerline{$^{64}$University of Mississippi, University, Mississippi 38677,  
                   USA}                                                       
\centerline{$^{65}$University of Nebraska, Lincoln, Nebraska 68588, USA}      
\centerline{$^{66}$Princeton University, Princeton, New Jersey 08544, USA}    
\centerline{$^{67}$Columbia University, New York, New York 10027, USA}        
\centerline{$^{68}$University of Rochester, Rochester, New York 14627, USA}   
\centerline{$^{69}$State University of New York, Stony Brook,                 
                   New York 11794, USA}                                       
\centerline{$^{70}$Brookhaven National Laboratory, Upton, New York 11973, USA}
\centerline{$^{71}$Langston University, Langston, Oklahoma 73050, USA}        
\centerline{$^{72}$University of Oklahoma, Norman, Oklahoma 73019, USA}       
\centerline{$^{73}$Brown University, Providence, Rhode Island 02912, USA}     
\centerline{$^{74}$University of Texas, Arlington, Texas 76019, USA}          
\centerline{$^{75}$Southern Methodist University, Dallas, Texas 75275, USA}   
\centerline{$^{76}$Rice University, Houston, Texas 77005, USA}                
\centerline{$^{77}$University of Virginia, Charlottesville, Virginia 22901,   
                   USA}                                                       
\centerline{$^{78}$University of Washington, Seattle, Washington 98195, USA}  
}                                                                             

\date{May 9, 2005}

\begin{abstract}
We report the first direct search for the Kaluza-Klein (KK) modes of Randall-Sundrum gravitons using dielectron, dimuon, and diphoton events observed with the D\O\ detector operating at the Fermilab Tevatron $p\bar p$ Collider at $\sqrt{s} = 1.96$ TeV. No evidence for resonant production of gravitons has been found in the data corresponding to an integrated luminosity of $\approx 260$~pb$^{-1}$.  Lower limits on the mass of the first KK mode at the 95\% C.L. have been set between 250 and 785 GeV, depending on its coupling to SM particles.
\end{abstract}

\maketitle 
 
Phenomenological models inspired by string theory in which there exist additional spatial dimensions have recently been proposed to remedy some of the defects in the standard model (SM). These models may solve the hierarchy problem, allow for low-energy gauge coupling unification, and address the issues of flavor and CP-violation.

The Randall-Sundrum (RS) model~\cite{RS} of extra dimensions (ED) offers a rigorous solution to a pressing problem of the SM~--- an apparent large hierarchy between the Planck scale at which gravity is expected to become strong ($M_{\rm Pl} \sim 10^{16}$~TeV) and the electroweak symmetry breaking scale ($M_{\rm EW} \sim 1$~TeV). This is achieved through the geometry of a slice of the 5-dimensional Anti-deSitter space-time ($AdS_5$), with a curved metric $ds^2 = \exp(-2kR|\varphi|)\eta_{\mu\nu}dx^\mu dx^\nu - R^2 d\varphi^2$, where $0 \le |\varphi| \le \pi$ is the coordinate along the single ED of radius $R$, $k$ is the curvature of the $AdS_5$ space (the warp factor), $x^\mu$ are the convential (3+1)-space-time coordinates, and $\eta^{\mu\nu}$ is the metric tensor of the Minkowski space-time. A ``hidden'' (3+1)-dimensional brane (Planck brane) is placed at $\varphi = 0$ and the second brane (SM brane) is located at $\varphi = \pi$. Gravity originates on the Planck brane and the graviton wave function is exponentially suppressed away from the brane along the ED due to the warp factor. Consequently, the $O(M_{\rm Pl})$ operators on the Planck brane yield low-energy effects on the SM brane with a typical scale of $\Lambda_\pi = \overline{M}_{\rm Pl}\exp(-k\pi R)$, where $\overline{M}_{\rm Pl} \equiv M_{\rm Pl}/\sqrt{8\pi}$ is the reduced Planck mass. Thus, the hierarchy problem is solved if $\Lambda_\pi \sim 1$~TeV, which can be achieved with little fine tuning by requiring $kR \approx 10$. This is a natural solution, as the only fundamental scale in this model is $\overline{M}_{\rm Pl}$ and $k \sim R^{-1} \sim \overline{M}_{\rm Pl}$. 

In the simplest RS model~\cite{RS,RSext}, the only particles propagating in the ED are gravitons. Consequently, they appear as a Kaluza-Klein (KK) tower of massive excitations from the point of view of the SM brane and can be resonantly produced in $p\bar p$ collisions. The masses and widths of the KK-excitations are related to the parameters of the RS model. The zeroth KK mode ($G^{(0)}$) remains massless and couples to the SM fields with gravitational strength, $1/M_{\rm Pl}$, while the excited modes couple with the strength of $1/\Lambda_\pi$. The excited modes can decay into fermion-antifermion or diboson pairs, leading to the characteristic resonance structure in their invariant mass spectrum. In this Letter we report on a search for the first excited mode of the KK graviton, \gKK, in the dielectron, dimuon, and diphoton final states. Since the graviton has spin 2, its decay products are either found in the $s$-wave (diphotons) or $p$-wave (dileptons). This leads to the branching fraction of the graviton decay in a single dilepton channel ($\ell\ell$) to be half that of diphotons.

Phenomenologically, it is convenient to express the two RS parameters $k$ and $R$ in terms of two direct observables: the mass of the first excited mode of the graviton, $M_1$, and the dimensionless coupling to the SM fields, \kMPl, which governs both graviton production cross section $(\sim (\kMPl)^2)$ and the width of the graviton resonance. The theoretically preferred range for $M_1$ is between a few hundred GeV and a few TeV, while \kMPl\ is expected to be between 0.01 and 0.1. Larger values of the coupling would render the theory non-perturbative, while smaller would require an undesirably large amount of fine-tuning. Indirect limits on RS model parameters come from precision electroweak data (dominated by the $S$ parameter)~\cite{RSext}. There have been no dedicated searches for RS gravitons to date.

We used the D\O\ detector operating at the Fermilab Tevatron $p\bar p$ Collider at $\sqrt{s} = 1.96$ TeV with approximately 246 \ipb\ of data accumulated with dimuon triggers and 275 \ipb\ of data collected with single or di-electromagnetic (EM) triggers for this search. To maximize the reconstruction efficiency for dielectrons and diphotons, we did not use tracking confirmation and combined these two channels in a single, calorimeter-based, ``diEM'' channel. The detector, data acquisition system, and triggering are detailed elsewhere~\cite{D0}.

Offline, we required EM objects to have transverse energy $E_T > 25$~GeV, be isolated in the calorimeter and tracker, have significant fraction of their energy deposited in the EM calorimeter, and have their EM shower shape consistent with that expected for an electron. We accepted EM objects in the central ($|\eta_d| < 1.1$)~\cite{eta} and forward ($1.5 < |\eta_d| < 2.4$) regions of the calorimeter, but required at least one of them to be central.

The overall efficiency per electron was determined using $Z \to ee$ events, and is $(91 \pm 2)$\% in the central and $(82 \pm 2)$\% in the forward regions. The efficiency is uniform in $E_T$ and $\eta_d$, with the exception of the region close to the boundaries between the central and forward calorimeters, $1.0 < |\eta_d| < 1.1$, where it drops by a factor of two. Monte Carlo (MC) simulations show that the efficiency per photon is 5\% lower than that per electron. An additional inefficiency of 7\% per event arises from the trigger, EM objects lost in azimuthal cracks between the central calorimeter modules or overlaps with jets in the events.

\begin{figure*}[htb]
\begin{center}
\includegraphics[height=2.3in]{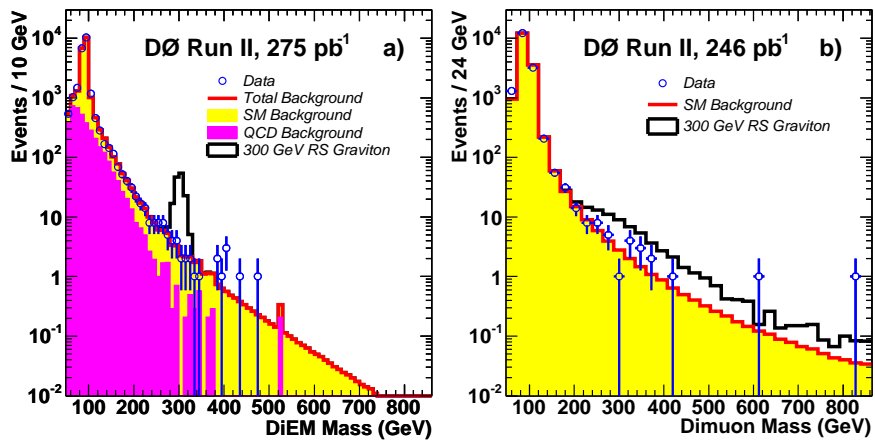}
\caption{Invariant mass spectrum in the a) diEM and b) dimuon channels. The points with error bars are data and the solid line is the overall background (dark shading in a) represents instrumental background). Also shown with an open histogram is the signal from an RS graviton with mass $M_1 = 300$~GeV and coupling \kMPl\ = 0.05.}
\label{fig:mass}
\vspace*{-0.3in}
\end{center}
\end{figure*}

Muons were identified in the muon spectrometer (covering $|\eta_d| \ltsim 2.0$) and were required to have a matching track in the central tracking detector, transverse momentum $p_T > 15$~GeV, be isolated, and pass additional hit and track quality requirements. Since the muon momentum resolution degrades rapidly at high $p_T$, high-mass dimuon events sometimes have the momentum of one of the muons misreconstructed. To remedy this and reduce  non-Gaussian tails in the invariant mass resolution, we assigned both muons the same value of transverse momentum, based on the weighted average (in 1/$p_T$) of their individual $p_T$'s. This results in $\approx 30\%$ decrease in the RMS of the invariant mass distribution at the cost of a modest ($\approx 1\%$) decrease in the invariant mass resolution. To reduce cosmic ray background, muon arrival times in the muon detector were required to be consistent with that for particles originating from beam collisions. The two muons in the event were not required to have opposite signs, as the sign determination efficiency degrades fast at high $p_T$. The overall selection efficiency per muon is $(80 \pm 4)$\%, as determined using $Z \to \mu\mu$ events. 

The above requirements result in 22,786 (17,128) diEM (dimuon) events used in the analysis. 
The main background to the RS graviton signal is Drell-Yan (DY) production in the dielectron and dimuon decay channels and direct diphoton production in the diphoton channel. These backgrounds were estimated using the leading order (LO) MC generator of Ref.~\cite{KCGL}, augmented with a parametric simulation of the D\O\ detector~\cite{D0ED}. The simulation accounts for the calorimeter and tracker resolutions, primary vertex position, detector acceptance, and $p_T$ of the $\ell\ell$ or $\gamma\gamma$ system. The CTEQ5L~\cite{CTEQ} set of parton distribution functions (PDF) was used in the simulations. 

In the diEM channel, an additional instrumental background arises from QCD multijet and direct photon events, with one or more jets reconstructed as EM objects. This background was estimated from the data, by inverting the shower shape quality requirement, with the absolute normalization obtained in the $Z$-boson mass peak. The only other background is due to $\tau\tau$ production with $\tau$'s decaying via the electron or muon channel and is negligible at invariant masses above the $Z$-peak where the search is performed. Figure~\ref{fig:mass} shows the invariant mass spectrum in the diEM and dimuon channels and demostrates good agreement between the data and expected background.

The RS graviton signal was simulated with the {\sc pythia}~\cite{PYTHIA} MC event generator with the CTEQ5L PDF, followed by the parametric simulation of the D\O\ detector. The LO {\sc pythia} cross section was scaled by a constant $K$-factor of 1.34 to account for next-to-LO (NLO) effects, recently calculated~\cite{VanNeervenED} for graviton exchange and shown to be similar to those for SM DY production. We set limits on the ratio of the graviton production cross section and the next-to-NLO (NNLO) $p\bar p \to Z \to ee$ cross section of $254 \pm 10$ pb~\cite{VanNeerven}. Since the $Z$-peak is found in the candidate sample, this approach allows for in situ calibration and reduces the overall systematic uncertainty. We quote the limits on production of gravitons in terms of the absolute cross section, which is obtained by multiplying the limits on the ratio by 254~pb.

A simulated signal is shown in Fig.~\ref{fig:mass} for $M_1 = 300$ GeV and $\kMPl = 0.05$.
Since the muon momentum was measured in the tracker, while the EM energy was determined from the calorimeter, the difference in resolutions for the two detectors explains that the mass resolution in these two channels is so different. We used a conservative estimate of the muon momentum and EM energy smearing parameters in the detector response simulation by attributing the measured width of the $Z$ boson to the constant resolution term, which dominates at high masses. This choice leads to a somewhat broader than expected reconstructed signal and to conservative limits on signal cross section. 

\begin{table*}[t]
\caption{Counting experiments and 95\% C.L. upper limits (in fb) on $\sigma(p\bar p \to \gKK \to \ell\ell)$. All masses are expressed in GeV.}
\label{table:counting}
\centering
\begin{tabular}{ccrclccccrclcccc}
\hline
\hline
Graviton & \multicolumn{7}{c}{DiEM Channel} & \multicolumn{7}{c}{Dimuon Channel} & Combined \\
Mass & ~~Window~~ & \multicolumn{3}{c}{Background} & Data & Limit & Sensitivity &
Window & \multicolumn{3}{c}{Background} & Data & Limit & Sensitivity & Limit \\
\hline
200 & 190--210 & $ 51.5 $&$ \pm $&$5.2$ & 53 & 70.2 & 68.2 & $>160$ & $90.1 $&$\pm $&$11.7 $ & 96 & 437 & 388 & 70.8 \\ 
220 & 210--230 & $30.7 $&$\pm $&$3.2 $ & 31 & 51.6 & 52.7 \\ 
240 & 230--250 & $17.8 $&$\pm $&$1.9 $ & 16 & 34.8 & 41.8 \\ 
250 & 240--260 & $ 14.1 $&$ \pm $&$1.5$ & 16 & 43.3 & 38.1 & $>200$ & $42.1 $&$\pm $&$5.5 $ & 46 & 256 & 224 & 43.9 \\ 
270 & 250--290 & $20.7 $&$\pm $&$2.2 $ & 25 & 46.7 & 36.2 \\ 
300 & 280--320 & $ 11.1 $&$ \pm $&$1.1$ & 12 & 28.9 & 27.4 & $>230$ & $26.2 $&$\pm $&$3.4 $ & 28 & 178 & 165 & 29.0 \\ 
320 & 300--340 & $8.27 $&$\pm $&$0.89 $ & 7 & 20.6 & 24.9 \\ 
350 & 330--370 & $ 5.80 $&$ \pm $&$0.73$ & 2 & 12.3 & 22.0 & $>250$ & $19.4 $&$\pm $&$2.5 $ & 24 & 186 & 141 & 13.0 \\ 
370 & 350--390 & $4.06 $&$\pm $&$0.51 $ & 2 & 13.1 & 19.3 \\ 
400 & 380--420 & $ 2.40 $&$ \pm $&$0.33$ & 6 & 30.5 & 16.7 & $>270$ & $14.7 $&$\pm $&$1.9 $ & 17 & 144 & 124 & 30.7 \\ 
450 & 420--480 & $ 1.92 $&$ \pm $&$0.30$ & 2 & 14.5 & 14.6 & $>280$ & $13.1 $&$\pm $&$1.7 $ & 17 & 152 & 113 & 15.4 \\ 
500 & 450--550 & $ 2.02 $&$ \pm $&$0.31$ & 1 & 10.8 & 14.2 & $>290$ & $11.8 $&$\pm $&$1.5 $ & 13 & 113 & 105 & 11.0 \\ 
550 & 500--600 & $ 1.20 $&$ \pm $&$0.27$ & 0 & 8.4 & 12.4 & $>300$ & $10.2 $&$\pm $&$1.3 $ & 13 & 123 & 96.9 & 8.9 \\ 
600 & 540--660 & $ 0.67 $&$ \pm $&$0.26$ & 0 & 8.3 & 10.8 & $>300$ & $10.2 $&$\pm $&$1.3 $ & 13 & 123 & 96.6 & 8.8 \\ 
650 & 590--710 & $ 0.38 $&$ \pm $&$0.25$ & 0 & 8.3 & 9.8 & $>300$ & $10.2 $&$\pm $&$1.3 $ & 13 & 117 & 92.4 & 8.8 \\ 
700 & 620--780 & $ 0.30 $&$ \pm $&$0.25$ & 0 & 8.2 & 9.5 & $>300$ & $10.2 $&$\pm $&$1.3 $ & 13 & 117 & 91.8 & 8.7 \\ 
750 & 660--840 & 0.20&$_{\footnotesize -}^{\footnotesize +}$&$_{\footnotesize 0.20}^{\footnotesize 0.25}$ & 0 & 8.1 & 8.9 & $>300$ & $10.2 $&$\pm $&$1.3 $ & 13 & 113 & 89.0 & 8.5 \\ 
800 & 700--900 & 0.13&$_{\footnotesize -}^{\footnotesize +}$&$_{\footnotesize 0.13}^{\footnotesize 0.25}$ & 0 & 8.1 & 8.7 & $>300$ & $10.2 $&$\pm $&$1.3 $ & 13 & 115 & 90.3 & 8.6 \\ 
\hline
\hline
\vspace{-0.3in}
\end{tabular}
\end{table*}

To set limits on graviton production, we performed analyses in a series of overlapping windows corresponding to different graviton masses. The width and position of the windows were optimized to give the highest signal sensitivity via a modified method of Ref.~\cite{GLKM}, which takes into account Gaussian fluctuations of an exponentially falling background. For the diEM channel at high masses ($> 300$ GeV), the background is small so a symmetric window with the width set to six times the detector resolution was used to maximize the sensitivity. Since the muon momentum resolution effects on the invariant mass are very asymmetric and result in a long high-mass tail (see Fig.~\ref{fig:mass}), only the lower mass bound is used in the dimuon channel windows. Since the internal graviton width is negligible compared to the instrumental resolution in the range of $M_1$ and \kMPl we studied, the window size did not depend on \kMPl. The overall geometrical acceptance for the signal in the diEM channel varies between 45\% and 62\%, depending on the mass point. In the dimuon channel, the corresponding variation is between 55\% and 67\%.

The results of the counting experiments are listed in Table~I. As the number of events in each window is consistent with the expected background (the significance of an upward fluctuation in the diEM channel at 400 GeV is $<2$ standard deviations), we set limits on the graviton production cross section. The limits were set independently in the two channels using a Bayesian technique~\cite{Bayes} with a flat prior for the signal and systematic uncertainties on signal and background taken into account. In the diEM channel, the signal uncertainty is 9\%, dominated by the mass-dependence of the EM efficiency (5\%), acceptance calculation (5\%), and difference between the photon and electron efficiencies (5\%). In the dimuon channel, the signal uncertainty is 8\%, dominated by the acceptance uncertainty (7\%). The common source of systematics for both channels is the 4\% $Z$ boson NNLO cross section uncertainty. The SM background uncertainty is 9--12\% and dominated by the $K$-factor mass dependence (5\%), efficiency determination (7\% in the diEM and 5\% in the dimuon channels), modeling of the momentum smearing (6\%, dimuon channel), and the PDF dependence (5\%). The uncertainty on the instrumental background in the diEM channel is dominated by the low statistics of the background sample at high masses. 

\begin{figure}[t]
\begin{center}
\includegraphics[height=2.4in]{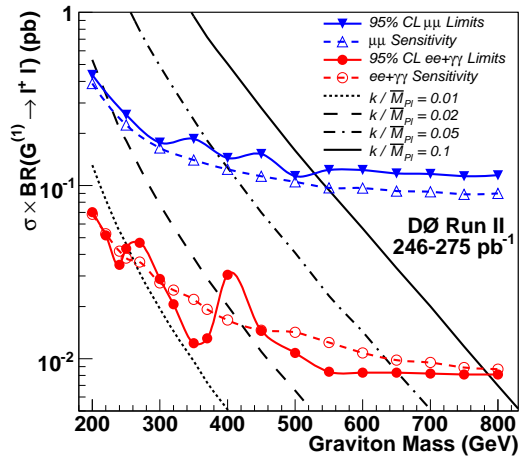}
\caption{The 95\% C.L. upper limits on $\sigma(p\bar p \to \gKK \to \ell\ell)$, as a function of the graviton mass. The upper (lower) solid line with points corresponds to the dimuon (diEM) channel. The dashed lines with points represent the expected limits. Also shown with a series of smooth lines the production cross sections for various values of \kMPl\ between 0.01 and 0.1.}
\label{fig:limits}
\vspace*{-0.3in}
\end{center}
\end{figure}

\begin{figure}[t]
\begin{center}
\includegraphics[height=2.4in]{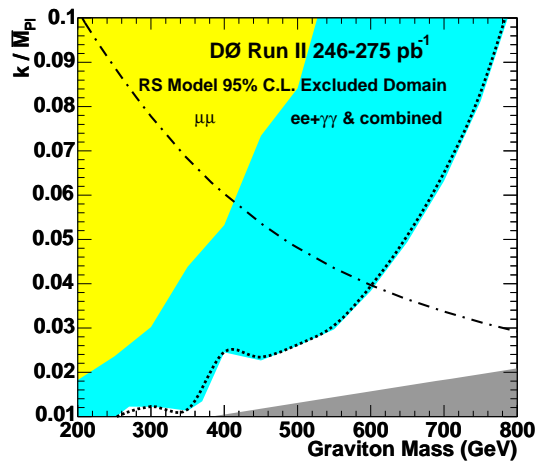}
\caption{95\% C.L. exclusion limits on the RS model parameters $M_1$ and \kMPl. The light-shaded area has been excluded in the dimuon channel; the medium-shaded area shows the extension of the limits obtained in the diEM channel; the dotted line corresponds to the combination of the two channels. The area below the dashed-dotted line is excluded from the precision electroweak data (see Ref.~\protect\cite{RSext}). The dark shaded area in the lower right-hand corner corresponds to $\Lambda_\pi > 10$ TeV, which requires a significant amount of fine-tuning.}
\vspace*{-0.3in}
\label{fig:exclusion}
\end{center}
\end{figure}

The 95\% C.L. upper limits on $\sigma(p\bar p \to \gKK \to \ell\ell)$ are listed in Table~I and shown in Fig.~\ref{fig:limits}. Also shown is the expected sensitivity of the search in each channel, defined as an average limit expected given the Poisson distribution of the background around its mean. We further combined the diEM and dimuon limits after taking into account common systematic uncertainties. The combined limits are very close to the diEM limits (and in fact are slightly less restrictive due to the overall small excess of observed events in the dimuon channel). We translate the limits on the cross section times branching fraction into limits on the RS model parameters $M_1$ and \kMPl, as shown in Fig.~\ref{fig:exclusion}. We did not include an uncertainty on the signal cross section related to the PDF and higher-order QCD effects. Assuming that it is similar to that for DY production ($\approx 10\%$) increases the cross section limits by 2.5\%. This translates into a negligible ($\approx 1\%$) fractional change in our limits on \kMPl\ for any graviton mass.

To conclude, we have performed the first dedicated search for Randall-Sundrum gravitons in the dielectron, dimuon, and diphoton channels using 246--275 \ipb\ of data collected by the D\O\ experiment in the Run II of the Fermilab Tevatron Collider. We see no evidence for resonant production of the first Kaluza-Klein mode of the graviton and set the most restrictive limits on the RS model parameters to date. Graviton masses up to 785 (250) GeV are excluded for \kMPl\ of 0.1 (0.01).

%
We thank the staffs at Fermilab and collaborating institutions, 
and acknowledge support from the 
DOE and NSF (USA),
CEA and CNRS/IN2P3 (France),
FASI, Rosatom and RFBR (Russia),
CAPES, CNPq, FAPERJ, FAPESP and FUNDUNESP (Brazil),
DAE and DST (India),
Colciencias (Colombia),
CONACyT (Mexico),
KRF (Korea),
CONICET and UBACyT (Argentina),
FOM (The Netherlands),
PPARC (United Kingdom),
MSMT (Czech Republic),
CRC Program, CFI, NSERC and WestGrid Project (Canada),
BMBF and DFG (Germany),
SFI (Ireland),
A.P.~Sloan Foundation,
Research Corporation,
Texas Advanced Research Program,
Alexander von Humboldt Foundation,
and the Marie Curie Program.

\raggedbottom


\begin{thebibliography}{11}
\bibitem{eta}
We use a polar coordinate system with the $z$-axis in the direction of the proton beam. The pseudorapidity in the detector frame $\eta_d$ is related to the polar angle $\theta$ measured with respect to the geometrical center of the detector as $\eta_d = -\ln(\tan\theta/2)$.
\end{thebibliography}
\end{document}